\documentclass[preprint,3p,twocolumn]{elsarticle}
 \usepackage{amsmath,amssymb,amsthm}
\usepackage{mathtools}
\usepackage{mathrsfs}
\usepackage{bm}
\usepackage{relsize}
\usepackage{graphicx}
\usepackage{color}
\usepackage[usenames,dvipsnames]{xcolor}
\usepackage{lineno}
\usepackage{hyperref}
\usepackage{cancel}
\usepackage{soul}
\usepackage[normalem]{ulem}

\colorlet{darkgreen}{green!50!black}
\colorlet{brightyellow}{yellow!75!red}
\colorlet{orange}{red!50!yellow}
\colorlet{darkblue}{blue!60!black}
\colorlet{darkred}{red!80!black}
\newcommand{\dd} {{\mathrm{d}}}
\newcommand{\ket}[1] {{\left.|#1\right>}}

\newcommand{\half}[1][1] {{\mathsmaller{\frac{#1}{2}}}}
\journal{arXiv}

\begin{document}

\begin{frontmatter}

\title{\textit{Ab Initio} Approach to the Non-Perturbative Scalar Yukawa Model}

\author[ISU]{Yang~Li}\corref{c1}
\ead{leeyoung@iastate.edu}

\author[LPI]{V.~A.~Karmanov}
\ead{karmanov@sci.lebedev.ru}

\author[ISU]{P.~Maris}
\ead{pmaris@iastate.edu}

\author[ISU]{J.~P.~Vary}
\ead{jvary@iastate.edu}

\cortext[c1]{Corresponding author}

\address[ISU]{Department of Physics and Astronomy, Iowa State University, Ames, Iowa, USA. 50011}
\address[LPI]{Lebedev Physical Institute, Leninsky Prospekt 53, 119991 Moscow, Russia }

\begin{abstract}
We report on the first non-perturbative calculation of the scalar 
Yukawa model in the single-nucleon sector up to four-body 
Fock sector truncation (one ``scalar nucleon'' and three ``scalar pions'').  The light-front 
Hamiltonian approach with a systematic non-perturbative renormalization is applied.
We study the $n$-body norms and the electromagnetic form
factor. We find that the one- and
two-body contributions dominate up to coupling $\alpha \approx 1.7$.
As we approach the coupling $\alpha \approx 2.2$, we discover that the
four-body contribution rises rapidly and overtakes the two- and
three-body contributions.
By comparing with lower sector truncations, we show that the form factor converges
 with respect to the Fock sector expansion.
\end{abstract}
\end{frontmatter}


\section{Introduction\label{sec 1}}

Solving quantum field theories in the non-perturbative regime is not
only a theoretical challenge but also essential to understand the
structure of hadrons from first principles. The light-front (LF)
Hamiltonian quantum field theory approach provides a natural framework
to tackle this issue \cite{Bakker2013.165,Perry1990.2959}.  A great
advantage of this approach is that it provides direct access to the
hadronic observables.
In the LF dynamics, the system is defined at a fixed LF
time $x^+ \equiv t + z$. The physical states are obtained by
diagonalizing the LF Hamiltonian operator.  The vacuum in LF
quantization is trivial. As a result, it is particularly convenient to
expand the physical states in the Fock space. For example, a physical
pion state can be written in terms of quarks ($q$), antiquarks ($\bar
q$) and gluons ($g$) as $\ket{\pi} = \ket{q \bar q} + \ket{q \bar q g}
+ \ket{q \bar q gg} + \cdots$.

In order to do practical calculations, the Fock space has to be
truncated. A natural choice, taking advantage of the LF dynamics, is
the Fock sector truncation, also known as the light-front Tamm-Dancoff
(LFTD) \cite{Perry1990.2959}.  A number of non-perturbative
renormalization schemes have been developed based on the LFTD
\cite{Perry1991.4051,Glazek1993.5863,Hiller1998.016006,Karmanov2008.085028}.
Thus we arrive at a few-body problem and predictions can be
systematically improved by including more Fock sectors. The LFTD
method is a non-perturbative approach in Minkowski space, which can be
compared with other non-perturbative methods, e.g, Lattice quantum
field theory in Euclidean space.
Of course, this approach only works if the Fock sector expansion
converges in the non-perturbative region.  In practice, one can
compare successive Fock sector truncations and check numerically
whether the relevant physical observables converge.
We will see that good convergence is achieved for the scalar Yukawa
model in a non-perturbative regime with a four-body Fock sector
truncation.  Similar results, though by a different method, were found
in Refs.~\cite{Hwang2004.413,Brodsky2006.1240} for the Wick-Cutkosky
model \cite{Wick1954.1124}.

We apply this approach to a scalar version of the Yukawa model 
that describes the pion-mediated nucleon-nucleon interaction. 
The Lagrangian density of the model reads
\begin{linenomath*}
\begin{multline} \label{eq 1}
 \mathscr{L} =
   \partial_\mu N^\dagger \partial^\mu N - m^2 |N|^2
+ \half \partial_\mu \pi \partial^\mu \pi - \half \mu^2_0 \pi^2 \\
+ g_0 |N|^2 \pi + \delta m^2 |N|^2,
\end{multline}
\end{linenomath*}
where $g_0$ is the bare coupling, $\delta m^2$ is the mass
counterterm of the field $N(x)$.  It is convenient to introduce a
dimensionless coupling constant 
\begin{linenomath*}
 \begin{equation*}
 \alpha = \frac{g^2}{16\pi m^2}.
 \end{equation*}
\end{linenomath*}
For the sake of brevity, we refer to the fundamental
degrees-of-freedoms (d.o.f.'s) $N(x)$ and $\pi(x)$ as ``scalar nucleon''
and ``scalar pion'' field respectively.
We also introduce a Pauli-Villars (PV) scalar pion (with mass $\mu_1$) to
regularize the ultraviolet (UV) divergence
\cite{Brodsky2001.114023}. Then, a sector dependent method known as
the Fock sector dependent renormalization (FSDR) developed in
Ref.~\cite{Karmanov2008.085028} is used to renormalize the theory. FSDR is
a systematic non-perturbative renormalization scheme based on the
covariant light-front dynamics (CLFD, see
Ref.~\cite{Carbonell1998.215} for a review) and Fock sector
expansion. It has shown great promise in the application to the Yukawa
model and QED \cite{Karmanov2010.056010,Karmanov2012.085006}.

The scalar Yukawa model is known to exhibit a vacuum instability
\cite{Baym1960.886}. It can be stabilized by either adding the quartic
terms $\frac{1}{4!}\pi^4$, $\frac{1}{2}|N|^4$ and
$\frac{1}{2}|N|^2\pi^2$ to the Lagrangian, or restricting the
nucleon-antinucleon d.o.f.~\cite{Franz2001.076008} The latter leads to
the exclusion of the pion self-energy correction, sometimes referred to as
the ``quenched approximation''. For the sake of simplicity, here we study
this restricted version of the theory.  Then the bare mass of the scalar pion 
becomes the physical mass, $\mu_0=\mu$.
It should be emphasized, though, that our formalism is capable of dealing with the 
(scalar) antinucleon d.o.f. The scalar nucleon and scalar pion d.o.f.'s generate non-perturbative
dynamics at large coupling sufficient for our purposes.

Previously, this model has been solved in the same approach up to
three-body truncation (one scalar nucleon, two scalar pions)
\cite{Karmanov2008.085028}.
The results from the two- and three-body truncations agree at small
couplings; yet they deviate in the large coupling region.  Therefore,
it is crucial to extend the non-perturbative calculation to higher
Fock sectors.
In this paper, we present the calculation of the four-body truncation
(one scalar nucleon, three scalar pions).  By comparing successive truncations, we
can examine the convergence of the Fock sector expansion. We presented
a preliminary version of this work in Ref.~\cite{Li:2014kfa}.

We first introduce our formalism in the next section. The LF
Hamiltonian field theory will be briefly mentioned and the
non-perturbative renormalization procedure will be explained.  Then a
set of coupled integral equations will be derived for the four-body
truncation.
In Sec.~\ref{sec 3}, we present the numerical results, including the
calculation of the electromagnetic form factor.  We conclude in
Sec.~\ref{sec 4}.

\section{Light-Front Hamiltonian Field Theory\label{sec 2}}

The LF Hamiltonian for the scalar Yukawa model is
\begin{linenomath*}
\begin{multline}
 \hat{P}^-  = \int  \mathrm{d}^3 x \,
\Big[
  \bm{\partial}_\perp N^\dagger \cdot \bm{\partial}_\perp N +  m^2 |N|^2  
 + \half \bm{\partial}_\perp \pi \cdot \bm{\partial}_\perp \pi  \\
 + \half \mu^2_0 \pi^2 - g_0 |N|^2 \pi - \delta m^2 |N|^2
\Big]_{x^+=0}. 
\end{multline}
\end{linenomath*}
The physical states can be obtained by solving the time-independent
Schr\"odinger equation
\begin{linenomath*}
\begin{equation}\label{schroedinger}
\hat{P}^- \ket{p} = \frac{\bm{p}^2_\perp+M^2}{p^+}\ket{p},
\end{equation}
\end{linenomath*}
where $\bm{p}_\perp$ and $p^+$ are the transverse and longitudinal
momentum, respectively.  Thanks to boost invariance in the LF
dynamics, we can take $\bm p_\perp = 0$ without loss of generality.

The system is solved in the single-nucleon sector. The state vector admits a Fock space expansion,
\begin{linenomath*}
\begin{multline}\label{eqn:Fock_expansion}
 \ket{p} = \sum_n \int D_n \, 
 \psi_n(\bm k_{1\perp}, x_1,\cdots \bm k_{n\perp},x_n;p^2) \\
\times \ket{\bm k_{1\perp}, x_1,\cdots \bm k_{n\perp},x_n},
\end{multline}
\end{linenomath*}
where $x_i \equiv \frac{k_i^+}{p^+}$, and
\begin{linenomath*}
\begin{multline*}
 D_n =  2 (2\pi)^3 \delta^{(2)} (\bm k_{1\perp}+\cdots \bm k_{n\perp}) \delta (x_1+\cdots x_n -1) 
\\
\times\prod_{i=1}^n\frac{\mathrm d^2 k_{i\perp} \mathrm d x_i}{(2\pi)^3 2x_i}.
\end{multline*}
\end{linenomath*}
The $n$-body Fock state $|\bm k_{1\perp}, x_1, \cdots, \bm k_{n\perp},
x_n\rangle$ consists $(n-1)$ scalar pions and 1 scalar nucleon. We use the last pair
$(\bm k_{n\perp}, x_n)$ to denote the momentum of the scalar
nucleon.  $\psi_n$, known as the LF wave function (LFWF), is a boost
invariant.  The LFWFs are normalized to unity, $ \sum_n I_n = 1$,
where
\begin{linenomath*}
\begin{equation}\label{normalization factors}
I_n = \mathsmaller{\frac{1}{(n-1)!}}\int D_n \Big|\psi_n(\bm k_{1\perp},x_1,\cdots \bm k_{n\perp},x_n;p^2) \Big|^2 
\end{equation}
\end{linenomath*}
is the probability that the system appears in the $n$-body Fock
sector. In the scalar Yukawa model, these quantities are regulator 
independent, in contrast to more realistic theories such as Yukawa and QED.
Note that $\psi_1 = \sqrt{I_1}$ is a constant.

\begin{figure*}[t!]
  \centering 
  \includegraphics[width=0.85\textwidth]{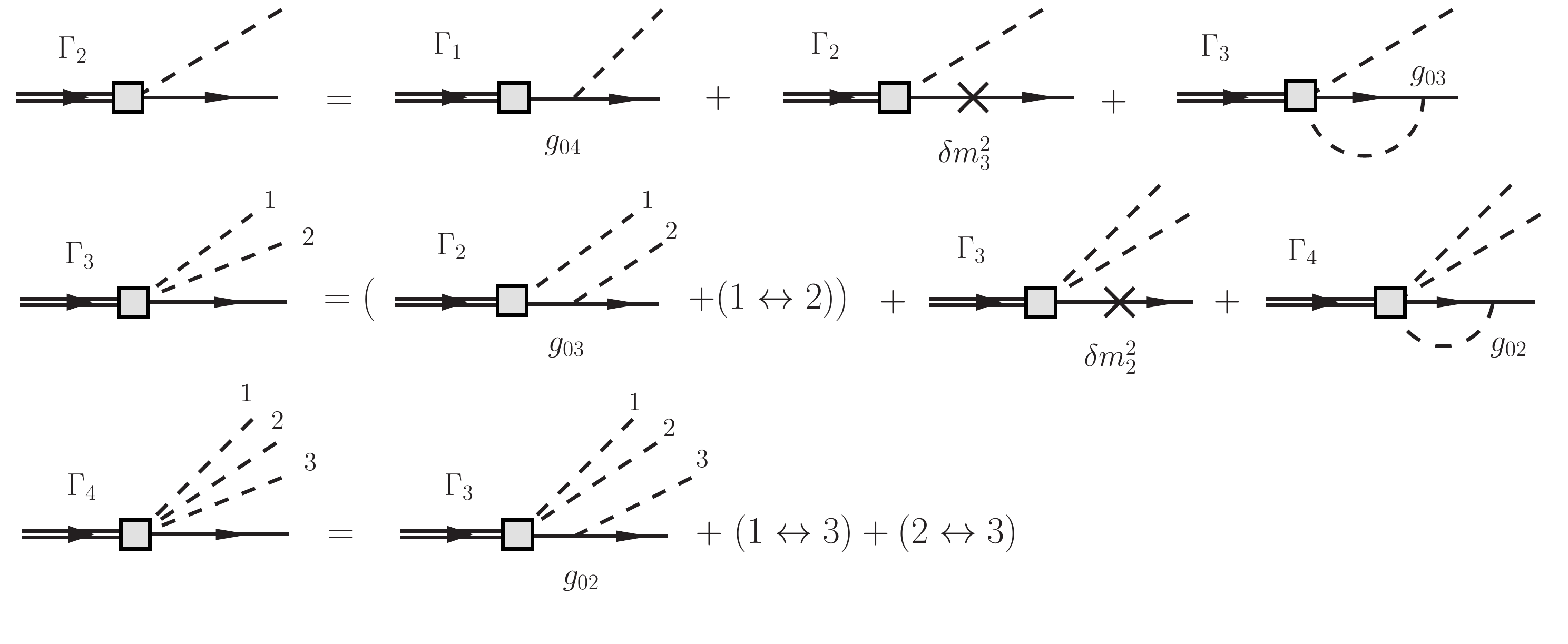}
  \caption{The diagrammatic representation of the system of equations
    in the four-body truncation.\label{fig:N4}}
\end{figure*}
\begin{figure*}[t!]
 \centering 
  \includegraphics[width=0.95\textwidth]{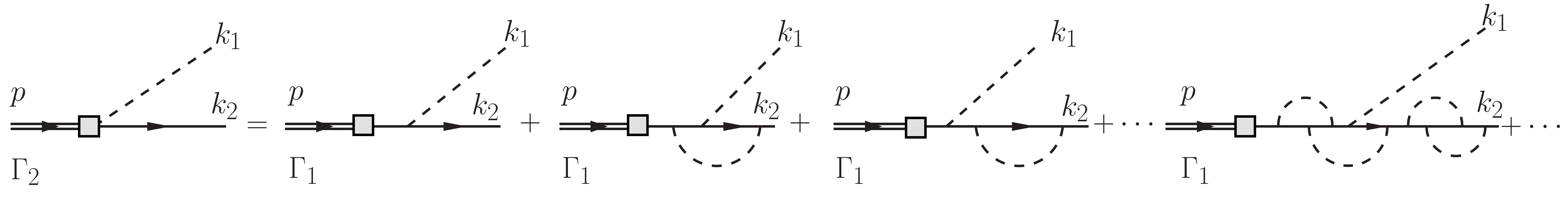} 
  \caption{ The perturbative expansion of the two-body vertex
    function.  The solid lines represent the scalar nucleons;
    the dashed lines represent the scalar pions; the double lines represent the
    dressed nucleons.\label{fig:vertex-function}}
\end{figure*}
It is convenient to introduce the $n$-body vertex functions,
\begin{linenomath*}
\begin{multline}
 \Gamma_n(\bm k_{1\perp},x_1,\cdots,\bm k_{n-1\perp}, x_{n-1};p^2) =\\
(s_{1,\cdots,n-1} - p^2) \psi_n (\bm k_{1\perp},x_1,\cdots, \bm k_{n\perp},x_n;p^2)
\end{multline}
\end{linenomath*}
for $n>1$ and $\Gamma_1 = (m^2-p^2)\psi_1$,
where 
\begin{linenomath*}
\begin{equation*}
\begin{split}
 s_{i_1,\cdots,i_{n-1}} \equiv & (k_{i_1}+\cdots k_{i_{n-1}} + k_n)^2 \\
= & \sum_{i=i_1}^{i_{n-1}} 
\frac{\bm k_{i\perp}^2+\mu^2_{j_i}}{x_i} + \frac{\bm k_{n\perp}^2+m^2}{x_n}
\end{split}
\end{equation*}
\end{linenomath*}
is the invariant mass squared of the Fock state, and $\mu_{j_i}$ ($j_i=0,1$) is the mass of the $i$-th scalar pion. 
We have suppressed $\bm k_{n\perp}$ and $x_n$ in $\Gamma_n$, by virtue of the momentum
conservations $\bm k_{1\perp}+\bm k_{2\perp}+\cdots \bm k_{n\perp} =
0$, $x_1+x_2+\cdots +x_n = 1$.
For simplicity we will also omit the dependence on $p^2$ in $\Gamma_n$
for the ground state $p^2 = m^2$.

Written in terms of the vertex functions $\Gamma$,
Eq.~(\ref{schroedinger}) can be represented diagrammatically using the
LF graphical rules \cite{Weinberg1966.1313, Karmanov1976.210} (see
Ref.~\cite{Carbonell1998.215} for a review).  Figure \ref{fig:N4} shows
the diagrams for the four-body truncation.

The two-body vertex function $\Gamma_2$ plays a particular role in
renormalization.  It comprises all radiative corrections allowed
by the Fock sector truncation, including the amputated vertex
$V_3(k_1,k_2,p)$ and the self-energy $\Sigma((p-k_1)^2)$
(see Fig.~\ref{fig:vertex-function}):
\begin{linenomath*}
\begin{equation}\label{two-body vertex}
 \Gamma_2(\bm k_{1\perp}, x_1;p^2) = Z((p-k_1)^2) V_3(k_1,k_2,p) \sqrt{I_1}, 
\end{equation}
\end{linenomath*}
where the function $Z(q^2) = \big(1 - \frac{\Sigma(q^2) - \Sigma(m^2)}{q^2 - m^2} \big)^{-1}$ 
is a generalization of the field strength renormalization 
constant $Z = \big(1-\frac{\partial}{\partial q^2}\Sigma(q^2)\big)^{-1}_{q^2=m^2} = I_1$.
Note the presence of the scalar pion spectator, which means that in the
$n$-body truncation, the self-energy correction in the expression for
$\Gamma_2$ is the $(n-1)$-body self-energy.

The dependence of renormalization constants on the Fock sector is a
general feature of the Fock sector expansion. We use $g_{0n}$ and
$\delta m_{n}^2$ to denote the bare coupling and the mass counterterm
from the $n$-body truncation, respectively.  According to the LSZ
reduction formula, the physical coupling $g = \mathcal T_{fi} =
\sqrt{Z} V^\star_3(k_1,k_2,p) \sqrt{I_1} $. Here ``$\star$'' means
that $V_3$ is evaluated at the renormalization point, the physical
mass shell $s_{1} = m^2 \Rightarrow \bm k^2_{1\perp} = -(1-x_1)\mu^2_0
-x_1^2m^2 \equiv \bm k^{\star 2}_{1\perp}$.  These relations provide
the on-shell renormalization condition
\cite{Hiller1998.016006,Karmanov2008.085028,Karmanov2010.056010},
\begin{linenomath*}
\begin{equation}\label{eqn:renormalization condition}
 \Gamma^{(n)}_2(\bm k_{1\perp}^\star,x_1;p^2 = m^2) = g \sqrt{Z^{(n-1)}}.  
\end{equation}
\end{linenomath*}
Here the Fock sector dependence is shown explicitly. For example,
$\Gamma_2^{(n)}$ represents the two-body vertex function found in the
$n$-body truncation.  Note that $\bm k^{\star 2}_{1\perp}$ is
negative, which means Eq.~(\ref{eqn:renormalization condition}) has to
be imposed through analytic continuation.

\begin{figure}[t]
  \centering 
  \includegraphics[width=0.4\textwidth]{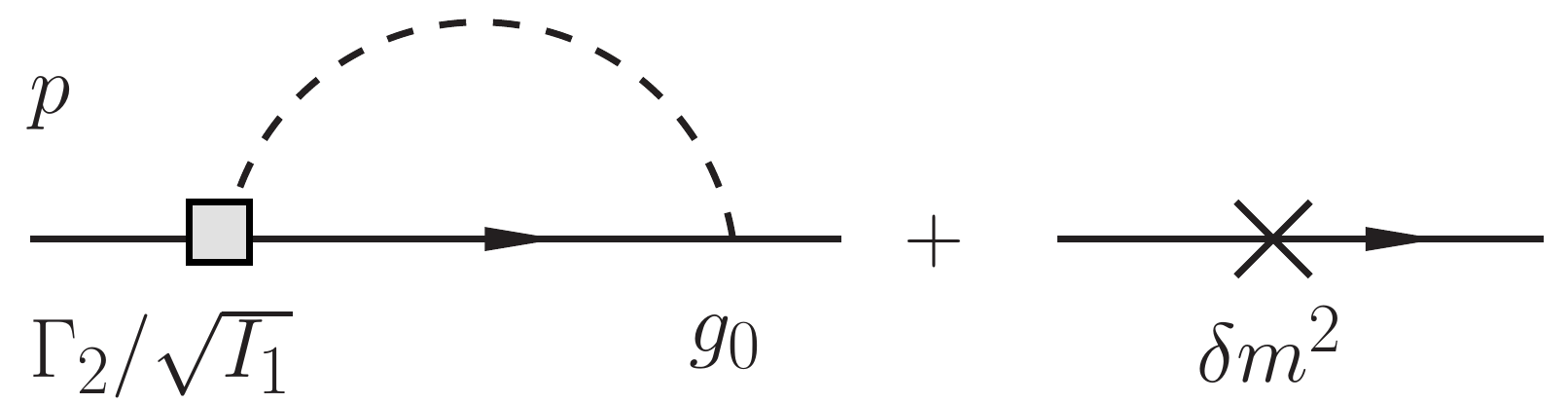}  
  \caption{The self-energy correction, loop correction $\Sigma$ plus
    mass counterterm $\delta m^2$, expressed in terms of the two-body
    vertex function $\Gamma_2$. Note the external lines are amputated. }
  \label{fig:selfenergy}
\end{figure}
The two-body vertex function $\Gamma_2$ also provides a
non-perturbative means to calculate the self-energy correction (see
Fig.~\ref{fig:selfenergy}). Following the LF graphical rules, the
self-energy in the $n$-body truncation is,
\begin{linenomath*}
\begin{multline}\label{eq6}
  \Sigma^{(n)}(p^2) = - \left( I^{(n)}_1 \right)^{-\half} \int 
\frac{\mathrm d^2 k_{1\perp}}{(2\pi)^3}\int_0^1 
\frac{\mathrm dx_1\,g_{0n}}{2x_1(1-x_1)} \\
\times \frac{\Gamma^{(n)}_2(\bm k_{1\perp},x_1;p^2)}{s_{1} - p^2}.
\end{multline}
\end{linenomath*}
Note that in our formalism the state vector in Eq.~(\ref{eqn:Fock_expansion}), rather than its 
one-body component, is normalized to unity. So according to the definition 
of the self-energy, the one-body LFWF $\psi_1 = \sqrt{I_1}$ is excluded 
from $\Gamma_2$ in the above expression.
Then the mass renormalization condition in the on-shell scheme implies
$\delta m^2_n = \Sigma^{(n)}(m^2)$.

As mentioned, the system of equations for $\Gamma_{2 \textrm{--} 4}$
resulted from truncating Eq.~(\ref{schroedinger}) to at most four-body
(one scalar nucleon and three scalar pions) are shown in Fig.~\ref{fig:N4}.  After
substituting $\Gamma_4$ into the second equation and applying the
renormalization condition Eq.~(\ref{eqn:renormalization condition}),
the system of equations becomes
\begin{linenomath*}
\begin{multline}\label{Gamma2}
  \Gamma^{j_1}_2(\bm k_{1\perp}, x_1) = {g}/{\sqrt{I_1^{(3)}}}
   +  \frac{\delta m_3^2\,\Gamma^{j_1}_2(\bm k_{1\perp},x_1)}{(1-x_1)(s_{1}-m^2)}  \\
  + \sum_{{j_2}=0}^1(-1)^{j_2} 
  \int \frac{\dd^2 k_{2\perp}}{(2\pi)^3} 
  \int\limits_0^{\mathclap{1-x_1}}\frac{\dd x_{2} \; g_{03}(\xi_{21}) }{2x_2(1-x_1-x_2)} \\
  \times  \bigg( \frac{\Gamma_3^{j_1j_2}(\bm k_{1\perp},x_1,\bm k_{2\perp},x_2)}{s_{12} - m^2} \\
 - \frac{\widetilde\Gamma^{0j_2}_3(\bm k^\star_{1\perp},x_1,\bm k_{2\perp},x_2)}{s_{12}^\star - m^2} \bigg)  
\end{multline}
\end{linenomath*}
\begin{linenomath*}
\begin{multline}\label{Gamma3}
   \Gamma^{j_1j_2}_3(\bm k_{1\perp},x_1,\bm k_{2\perp},x_2) = \\
   Z^{(2)}(q^2_{12})  \bigg[ \frac{g_{03}(\xi_{21})\,\Gamma^{j_1}_2(\bm k_{1\perp},x_1)}{(1-x_1)(s_{1}-m^2)} 
    + g_{02}^2 \sum_{j_3=0}^1(-1)^{j_3} \\
    \times \int \frac{\dd^2 k_{3\perp}}{(2\pi)^3}  \int\limits_0^{\mathclap{1-x_1-x_2}}  
\frac{\dd x_3}{2x_3(1-x_1-x_3)(1-x_1-x_2-x_3)} \\
   \times \frac{1}{s_{123} -m^2}  \frac{\Gamma^{j_1j_3}_3(\bm k_{1\perp},x_1,\bm k_{3\perp},x_3)}{(s_{13}-m^2)} \bigg] 
 + \big( 1 \leftrightarrow 2 \big) 
\end{multline}
\end{linenomath*}
where $\xi_{21} = x_2/(1-x_1)$, $s^\star_{12}= \frac{\bm k_{1\perp}^{\star 2} + \mu^2_0}{x_1} + \frac{\bm k^2_{2\perp} + \mu^2_{j_2}}{x_2} + 
\frac{(\bm k^\star_{1\perp}+\bm k_{2\perp})^2 + m^2}{1-x_1-x_2}$, $q^2_{12} = m^2 - (1-x_1-x_2)(s_{12}-m^2)$, and 
$Z^{(2)}$ comes from combining the two-body self-energy corrections. We have included the PV scalar pions ($j=1$) in 
the equations along with the ``physical'' pions ($j=0$). As mentioned, $g_{02}, \delta m_2^2, 
g_{03}, \delta m_3^2$ are sector dependent renormalization ``constants'' obtained from the 
two- and three-body truncations \cite{Karmanov2008.085028}. In fact, $g_{03}$ depends on the momentum
fraction $x$, which is a manifestation of the violation of the Lorentz symmetry by the Fock sector truncation 
\cite{Karmanov2010.056010}.
$\widetilde \Gamma_3$ is an auxiliary function that satisfies the integral equation, 
\begin{linenomath*}
\begin{multline}\label{tildeGamma3}
  \widetilde\Gamma^{0j_2}_3(\bm k^\star_{1\perp},x_1,\bm k_{2\perp},x_2) = \\
  Z^{(2)}(q^{\star 2}_{12}) \Bigg[ \frac{g_{03}(\xi_{12})\Gamma^{j_2}_2(\bm k_{2\perp},x_2)}{(1-x_2)(s_{2}-m^2)} 
+ g_{02}^2 \sum_{j_3=0}^1 (-1)^{j_3}  \\
 \times \int \frac{\dd^2 k_{3\perp}}{(2\pi)^3}  
 \int\limits_0^{\mathclap{1-x_1-x_2}}
\frac{\dd x_3 }{2x_3(1-x_1-x_3)(1-x_1-x_2-x_3)} \\
\times \frac{1}{s^\star_{123} -m^2}   
 \bigg( \frac{\widetilde\Gamma^{0j_3}_3(\bm k^\star_{1\perp},x_1,\bm k_{3\perp},x_3)}{s^\star_{13}-m^2} \\
+ \frac{\Gamma^{j_2j_3}_3(\bm k_{2\perp},x_2,\bm k_{3\perp},x_3)}{s_{23}-m^2} \bigg) 
\Bigg], 
\end{multline}
\end{linenomath*}
where 
$\xi_{12}=x_1/(1-x_2)$,
$q^{\star 2}_{12} = m^2 - (1-x_1-x_2)(s^\star_{12}-m^2)$,
$s_{123}^\star= \frac{\bm k^{\star 2}_{1\perp} + \mu^2_0}{x_1} + \frac{\bm k^2_{2\perp} + \mu^2_{j_2}}{x_2} 
+ \frac{\bm k_{3\perp}^2+\mu_{j_3}^2}{x_3} + \frac{(\bm k^\star_{1\perp}+\bm k_{2\perp}+\bm k_{3\perp})^2 + m^2}{1-x_1-x_2-x_3}$. 
Note that Eq.~(\ref{Gamma2}) can be eliminated by substituting $\Gamma_2$ into Eq.~(\ref{Gamma3}) and Eq.~(\ref{tildeGamma3}).

\section{Numerical Results\label{sec 3}}

\begin{figure}[ht]
 \centering 
 \includegraphics[width=0.48\textwidth]{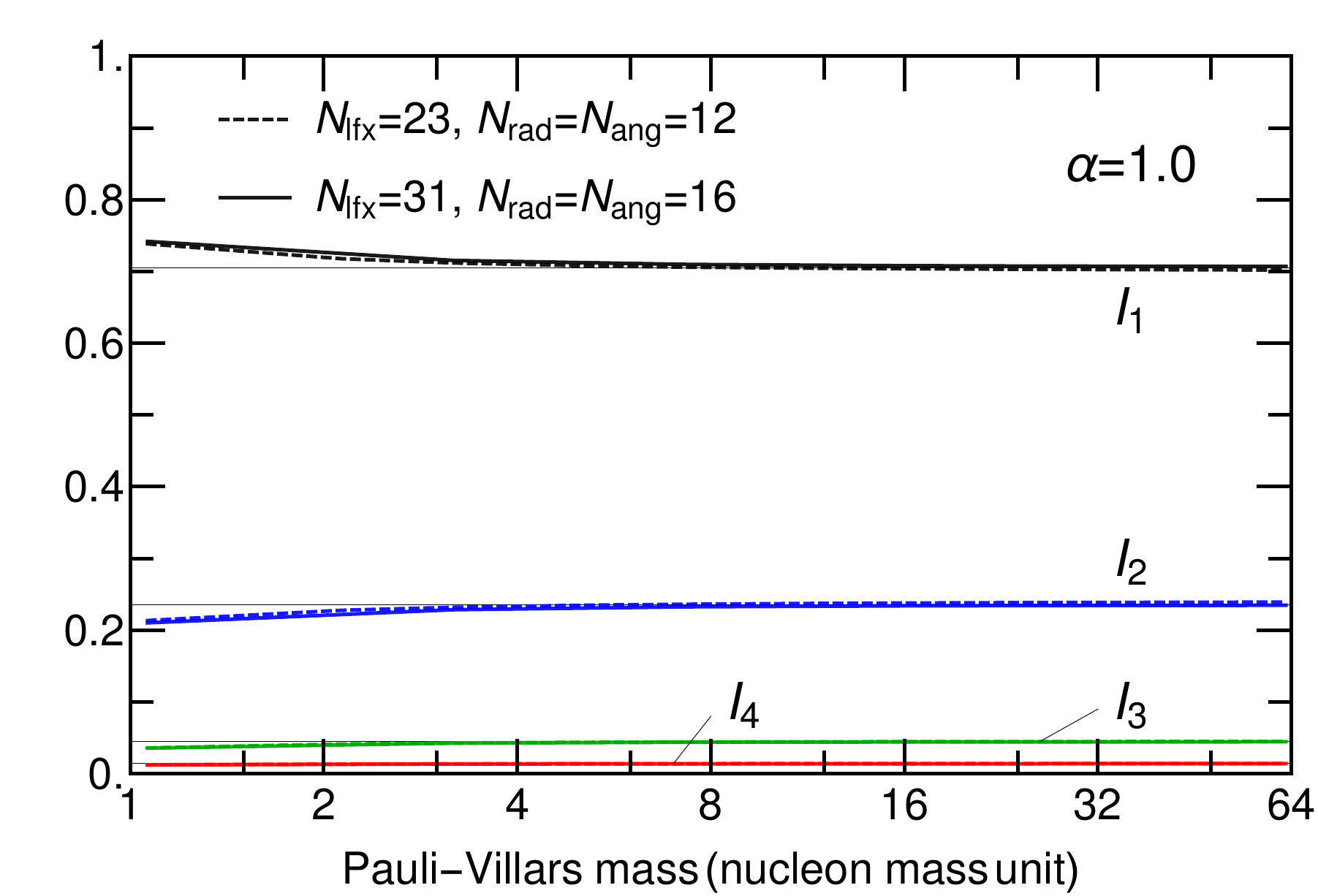}
 \includegraphics[width=0.48\textwidth]{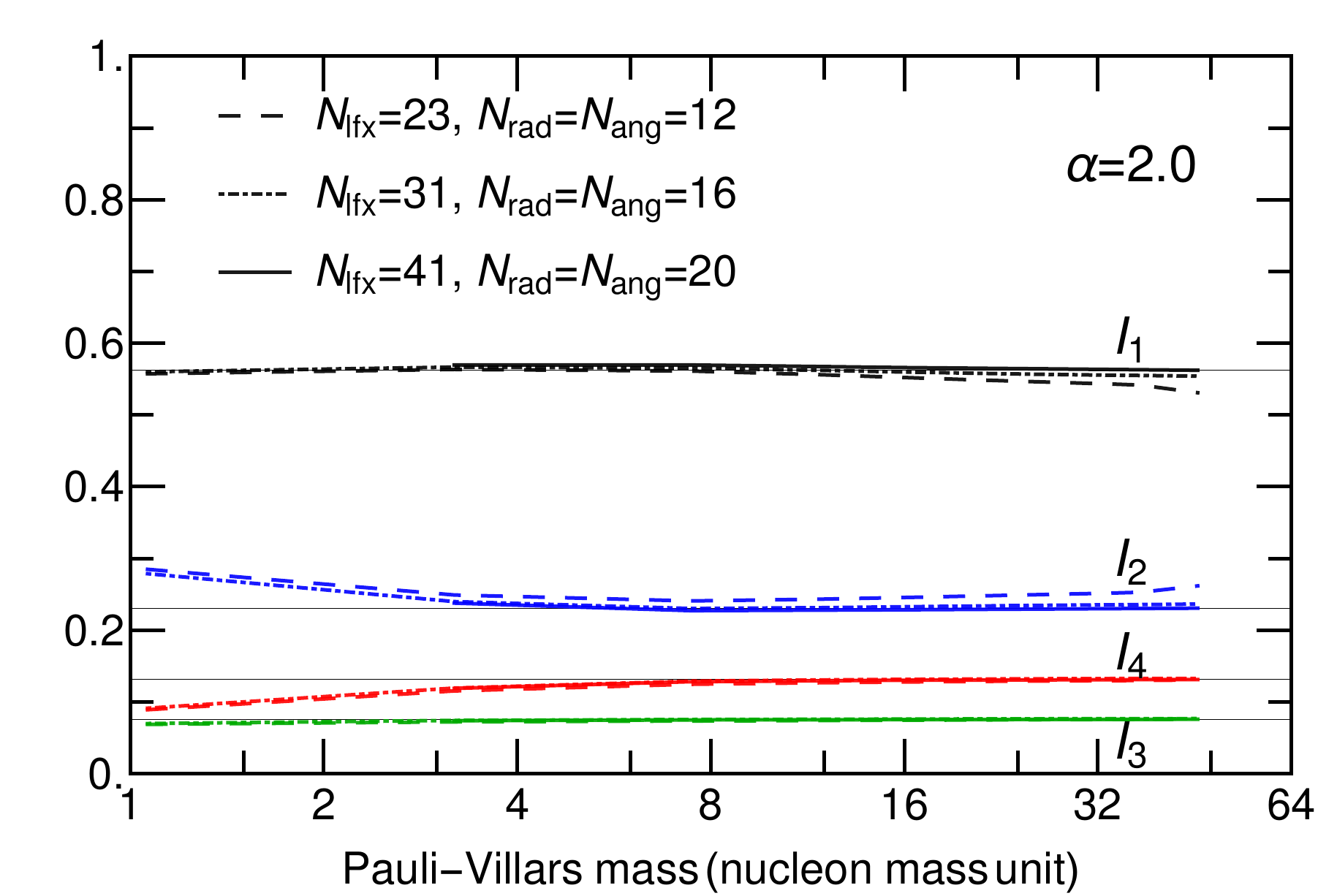}
 \caption{The Fock sector norms $I_{1-4}$ as a function of the PV mass
   $\mu_1$ for $\alpha=1.0$ (top) $\alpha = 2.0$ (bottom).  Results
   evaluated on different grids are shown. \label{fig:pv_conv}}
\end{figure}
We employ an iterative procedure to solve
Eqs.~(\ref{Gamma2}--\ref{tildeGamma3}).  The momenta are discretized
on chosen grids in the transverse radial and angular coordinates as
well as in the longitudinal coordinate, where the grid sizes are
controlled by the number of abscissas, $N_\text{rad}$, $N_\text{ang}$,
and $N_\text{lfx}$.  Then the integrals are approximated by the
Gauss-Legendre quadrature. We start with an initial guess of the
vertex functions and update them iteratively, until reaching
a pointwise absolute tolerance $\max\{ |\Delta \Gamma| \} < 10^{-4}$.
We solved the system at $m = 0.94\,\mathrm{GeV}, \mu_0 = 0.14\,\mathrm{GeV}$. 
The numerical results are obtained using Cray XE6 Hopper at NERSC.

Figure \ref{fig:pv_conv} plots the Fock sector normalization factors
$I_n$ (see Eq.~(\ref{normalization factors})) as a function of the PV
mass $\mu_1$ for two selected coupling constants. It shows that for
sufficiently large grids, $I_n$ converge as $\mu_1$ increases.
However, for a fixed grid, increasing $\mu_1$ would increase the
numerical error while decreasing the systematic error introduced by
the finite regulator, as larger $\mu_1$ requires more coverage in the
UV hence larger grid size.  A PV mass $\mu_1 = 15\;\mathrm{GeV}$
suffices for our purposes here.

There exist two critical couplings at $\alpha_c \approx 2.6$ and $\alpha'_c \approx 2.2$. In the two-body 
truncation, one finds the bare coupling, 
\begin{linenomath*}
\begin{equation*}
 \frac{1}{g^2} - \frac{1}{g_{02}^2} = 
\frac{1}{16\pi^2m^2} \left[ f\Big( \frac{\mu_0}{m} \Big) - f\Big( \frac{\mu_1}{m} \Big) \right], 
\end{equation*}
\end{linenomath*}
where
$ f(\lambda) = \int_0^1 \dd x\,{x(1-x)}/{((1-x)\lambda^2 +x^2)}$ and $f(\lambda \to \infty) = 0$.
If the physical coupling constant $\alpha > \alpha_c \equiv
\pi/f(\mu_0/m)$, the two-body bare coupling $g_{02}$ diverges at some
finite PV mass. Such a singularity in $g_{02}$ (known as the ``Landau
pole'' in a similar case in QED) propagates from the two-body truncation to the four-body
truncation via $g_{02}$ used in the FSDR.
At $\alpha = \alpha'_c$, the determinant of the Hamiltonian in the
three-body truncation crosses zero. 
Similarly, this singularity 
propagates from the three-body truncation to the four-body truncation and the iterative 
procedure in the four-body truncation diverges at $\alpha \gtrsim \alpha'_c$.

\begin{figure}
  \centering 
  \includegraphics[width=0.48\textwidth]{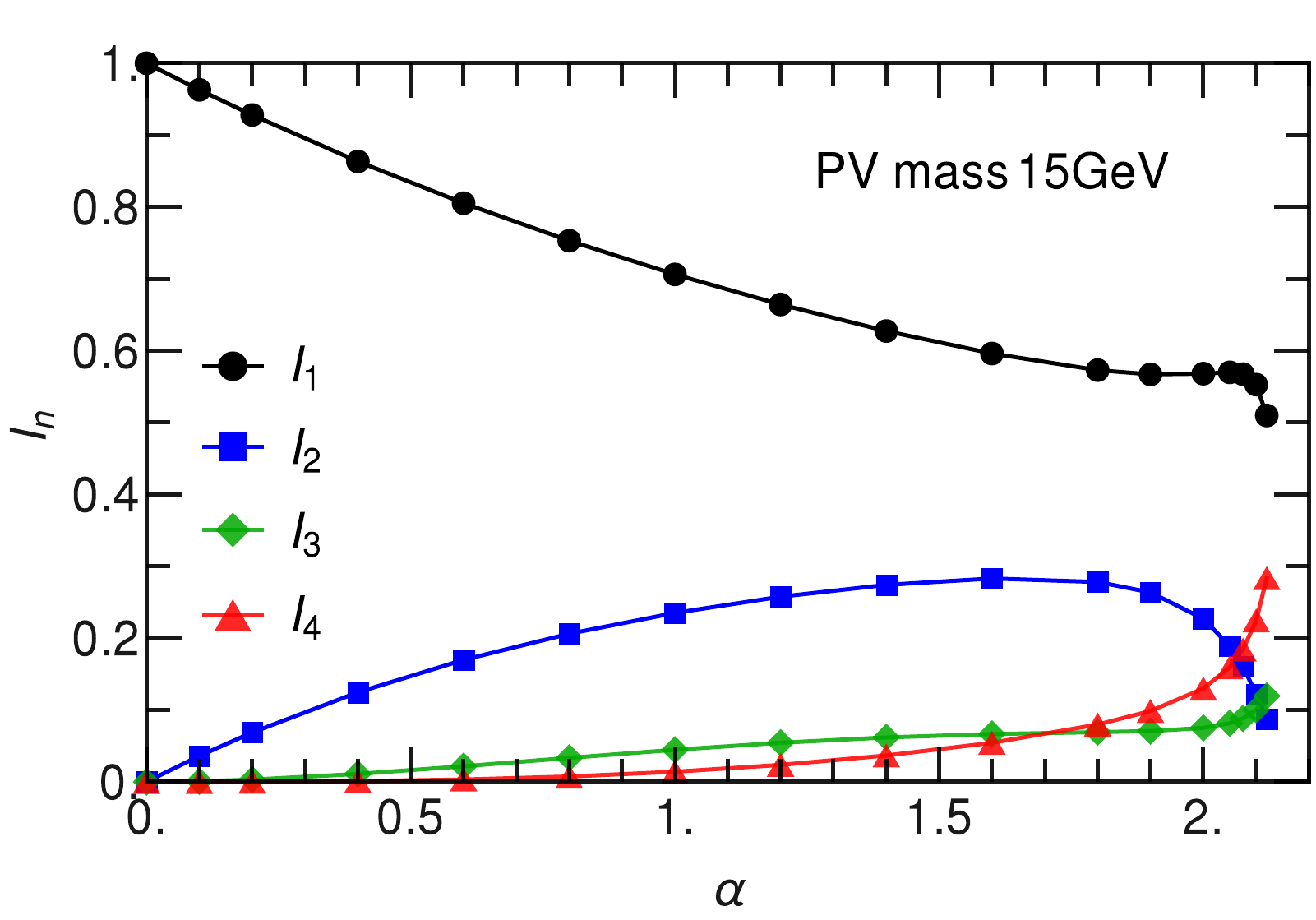} 
  \caption{
   The Fock sector norms $I_{1-4}$ as a function of the coupling 
   constant $\alpha$. Results are evaluated on the grid 
   $N_\mathrm{lfx}=41, N_\mathrm{rad}=N_\mathrm{ang}=20$, 
   with a PV mass $\mu_1=15\,\mathrm{GeV}$.}
   \label{fig:fock_norms} 
\end{figure}
Figure \ref{fig:fock_norms} shows the contribution of each Fock sector
in the four-body truncation for couplings up to $\alpha = 2.12$.  A
natural Fock sector hierarchy $I_1 > I_2 > I_3 > I_4$ can be observed,
up to $\alpha \approx 1.7$. Beyond $\alpha \approx 1.7$, $I_4$ exceeds $I_3$ and begins a steep 
climb with increasing $\alpha$. Meanwhile, $I_2$ turns over and starts to
fall. The net effect is that $I_4$ exceeds $I_2$ and $I_3$ at about $\alpha\approx 2.1$.
Clearly, as we approach $\alpha'_c$, dramatic changes in the $I_n$'s are emerging
and it appears that the Fock space expansion breaks down. Nevertheless, the lowest sectors 
$\ket{N}+\ket{N\pi}$ are observed to dominate the Fock space up to $\alpha \approx 2.0$, where
these two sectors constitute $80\%$ of the full norm.

\begin{figure}[t]
  \centering
  \includegraphics[width=0.48\textwidth]{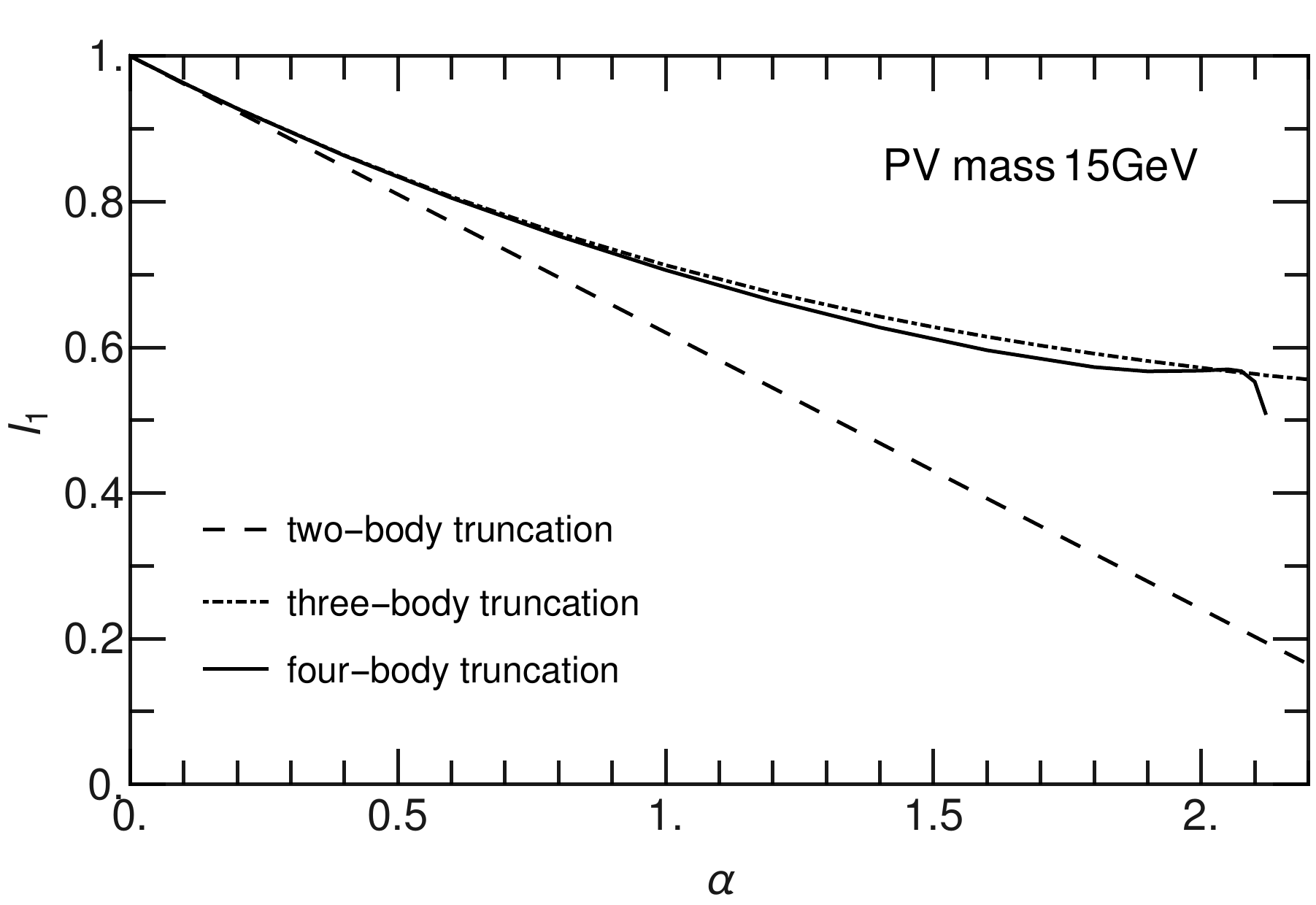} 
  \includegraphics[width=0.48\textwidth]{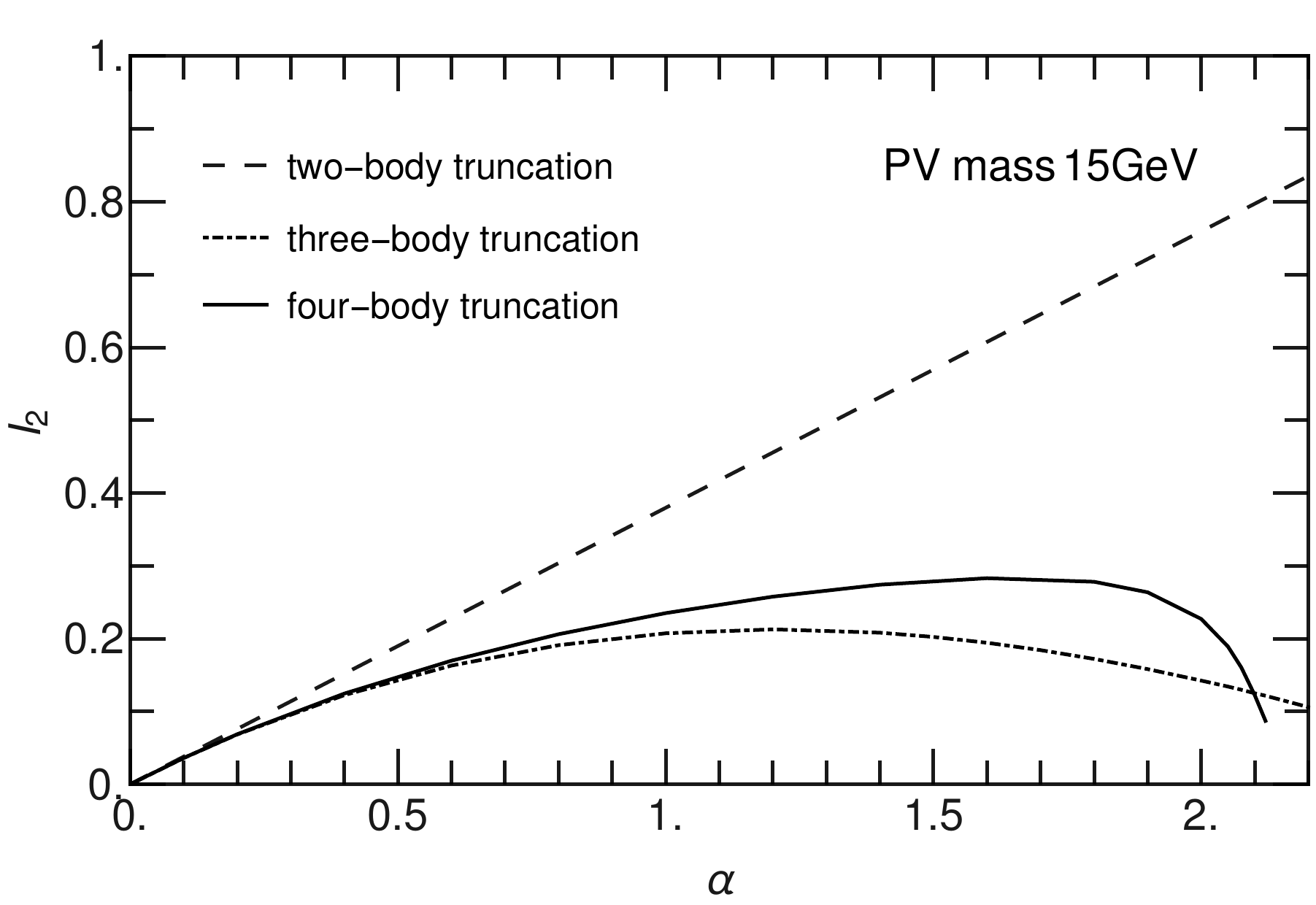} 
  \caption{Comparison of the Fock sector norms $I_1$ (top) and $I_2$
    (bottom) from successive two-, three- and four-body
    truncations.\label{fig:fock_norm_conv}}
\end{figure}
Figure \ref{fig:fock_norm_conv} compares the Fock sector norms from
the four-body truncation with their counterparts from the two- and
three-body truncations.  The result suggests a convergence as the
number of constituent bosons increases, especially for the coupling
below $\alpha \approx 1.0$.  Note that the one-body norm $I_1$ changes
little from the three-body truncation to the four-body truncation,
even around $\alpha\approx 1.7$.

\begin{figure}[ht]
 \centering 
 \includegraphics[width=0.48\textwidth]{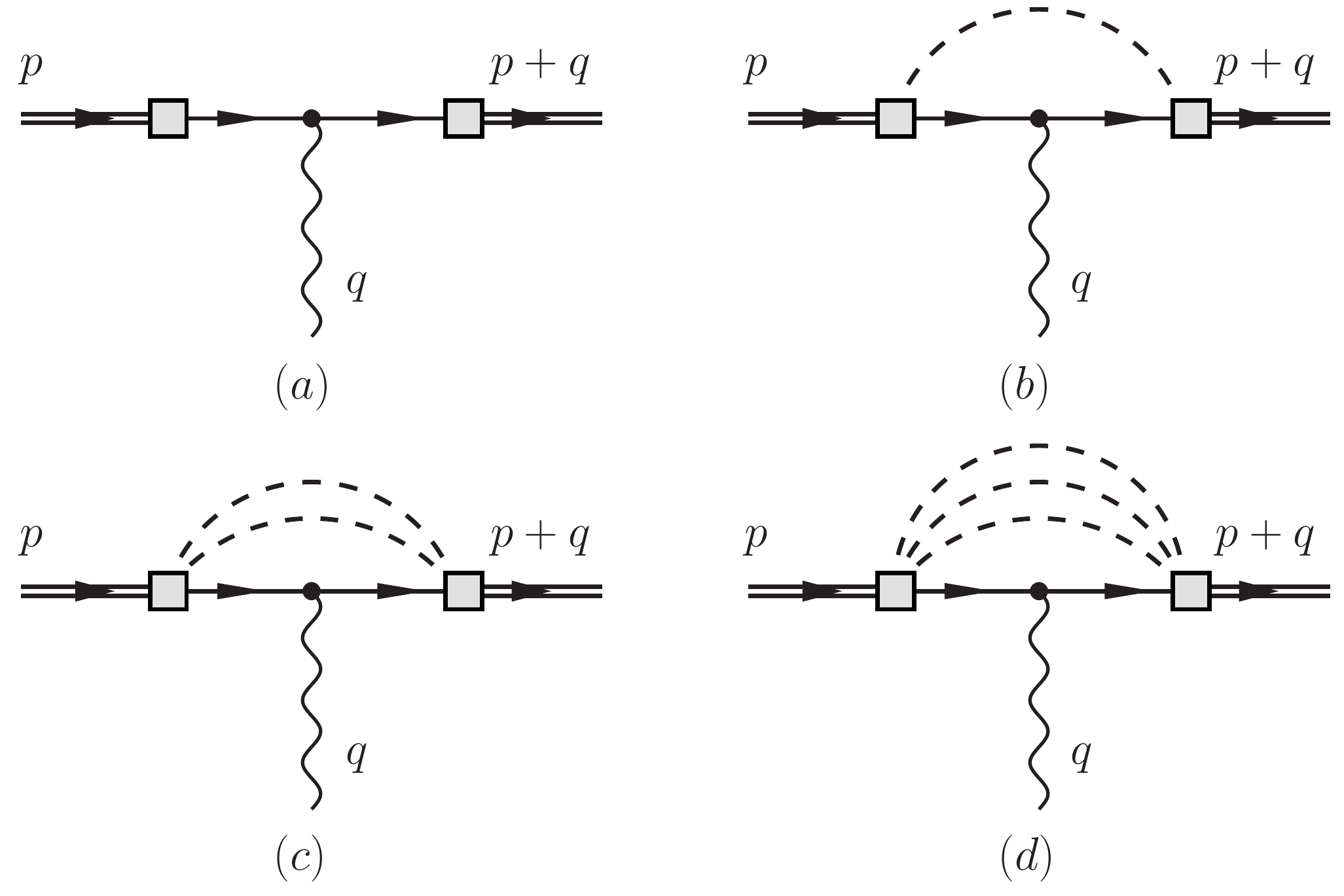}
 \caption{The various Fock sector contributions to the 
 electromagnetic form factor for the scalar nucleon (solid lines). 
 The dashed lines represent the internal scalar pions. 
 The wavy lines represent the external photons.
}
 \label{fig:formfactor_diagram}
\end{figure}
The obtained LFWFs are now available for computing physical
observables. Here we consider the elastic electromagnetic form factor
for photon coupling to the scalar nucleon,
which is obtained from the matrix element of the 
``+'' component of the current (see 
Fig.~\ref{fig:formfactor_diagram}),
\begin{linenomath*}
\begin{equation}
\langle p + q|J^+(0)|p \rangle =2p^+ F(Q^2),
\end{equation}
\end{linenomath*}
where $q^+=0$, $Q^2 = -q^2 = \bm q^2_\perp > 0$.
In LF dynamics, the form factor obtains the form \cite{Drell1970.181}:
\begin{linenomath*}
\begin{multline}
 F(Q^2) 
 = \sum_n \mathsmaller{\frac{1}{(n-1)!}} \int D_n 
 \psi^*_n(\bm k'_{1\perp}, x_1,\cdots, \bm k'_{n\perp},x_n) \\
 \times \psi_n(\bm k_{1\perp},x_1,\cdots,\bm k_{n\perp},x_n) 
\end{multline}
\end{linenomath*}
where $\bm k'_{i\perp} = \bm k_{i\perp} - x_i \bm q_\perp$, ($i=1,2,\cdots, n-1$), for the spectators and 
$\bm k'_{n\perp} = \bm k_{n\perp} + (1-x_n) \bm q_\perp$ for the struck parton.
\begin{figure}[ht]
  \centering 
  \includegraphics[width=0.48\textwidth]{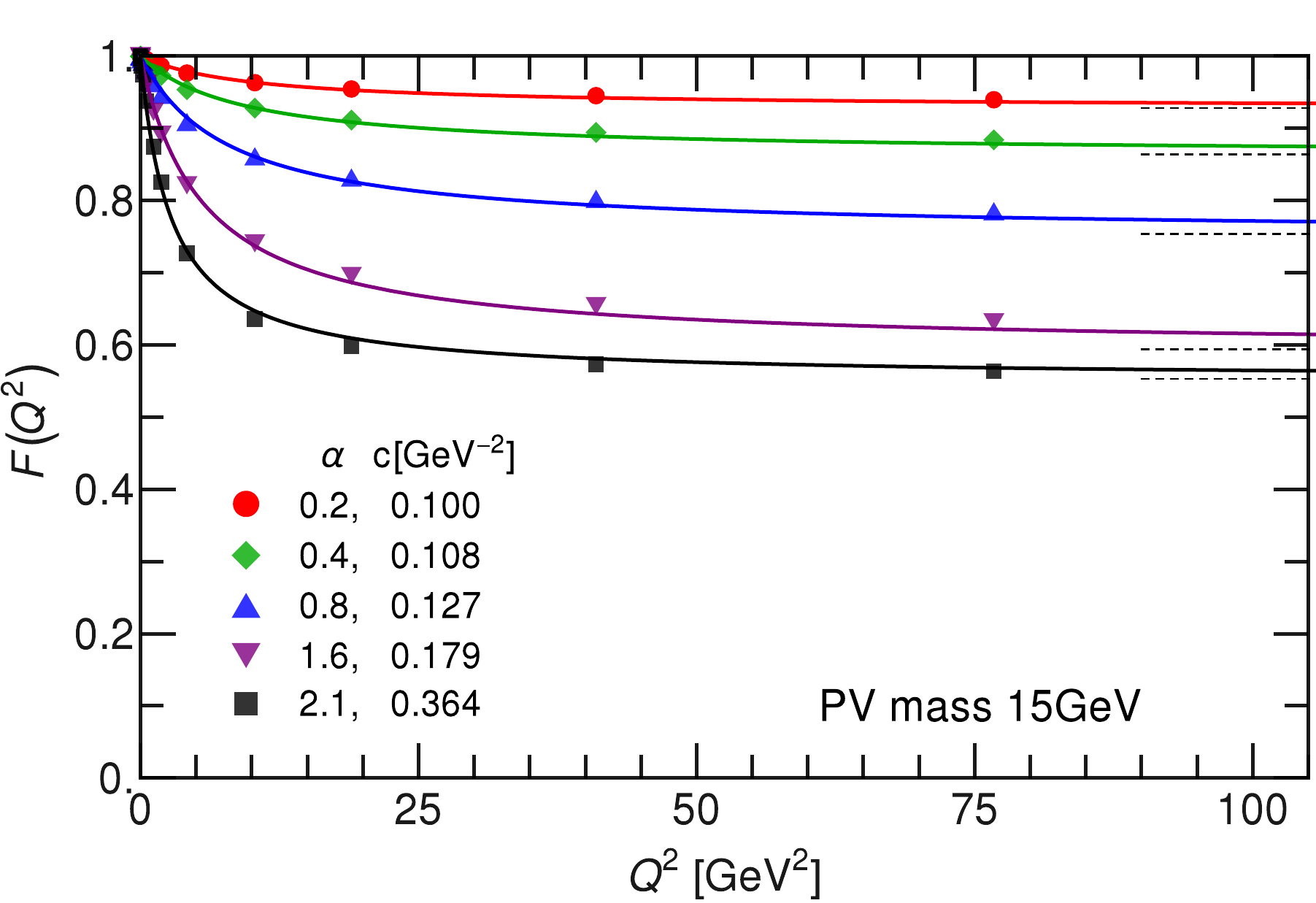}
  \caption{The elastic electromagnetic form factor $F(Q^2)$ in the
    four-body truncation for couplings $\alpha = 0.2$, $0.4$, $0.8$,
    $1.6$, and $2.1$.  The numerical results (symbols) are fitted by
    Eq.~({\ref{eqn:formfactorfit}}) (lines).\label{fig:formfactor}}
\end{figure}
Figure \ref{fig:formfactor} shows the form factor for some selected
couplings.  In the limit of $Q^2 \to 0$, $F(0) = 1$, consistent with
the charge conservation; in the limit of $Q^2 \to \infty$, $F(\infty)
= I_1$, representing a point-like charge. The form factors can be approximated by
\begin{linenomath*}
\begin{equation}\label{eqn:formfactorfit}
 F(Q^2) \approx \frac{1+c\,I_1 Q^2}{1+c\,Q^2}. 
\end{equation}
\end{linenomath*}

\begin{figure}[ht]
  \centering 
  \includegraphics[width=0.48\textwidth]{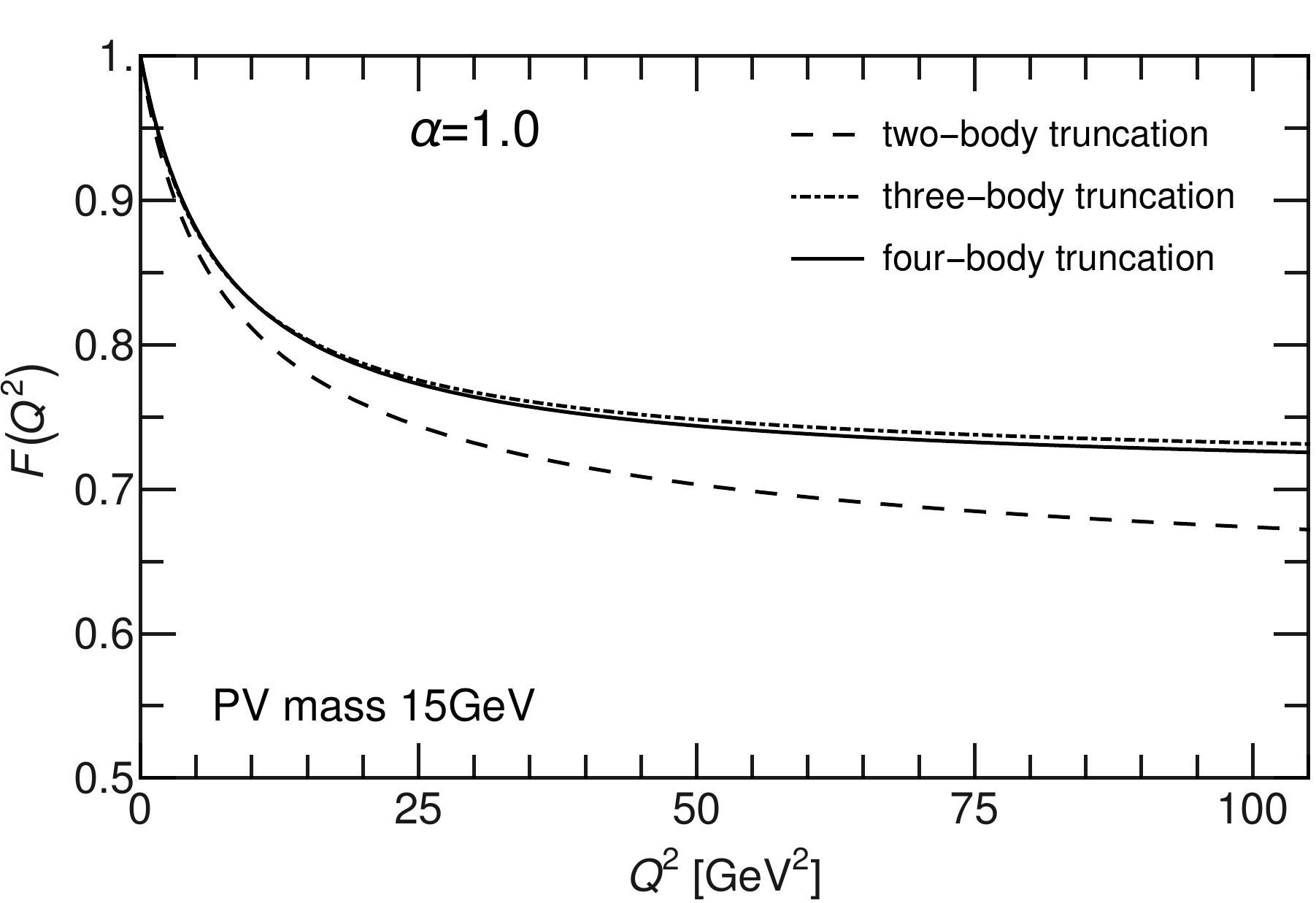}
  \includegraphics[width=0.48\textwidth]{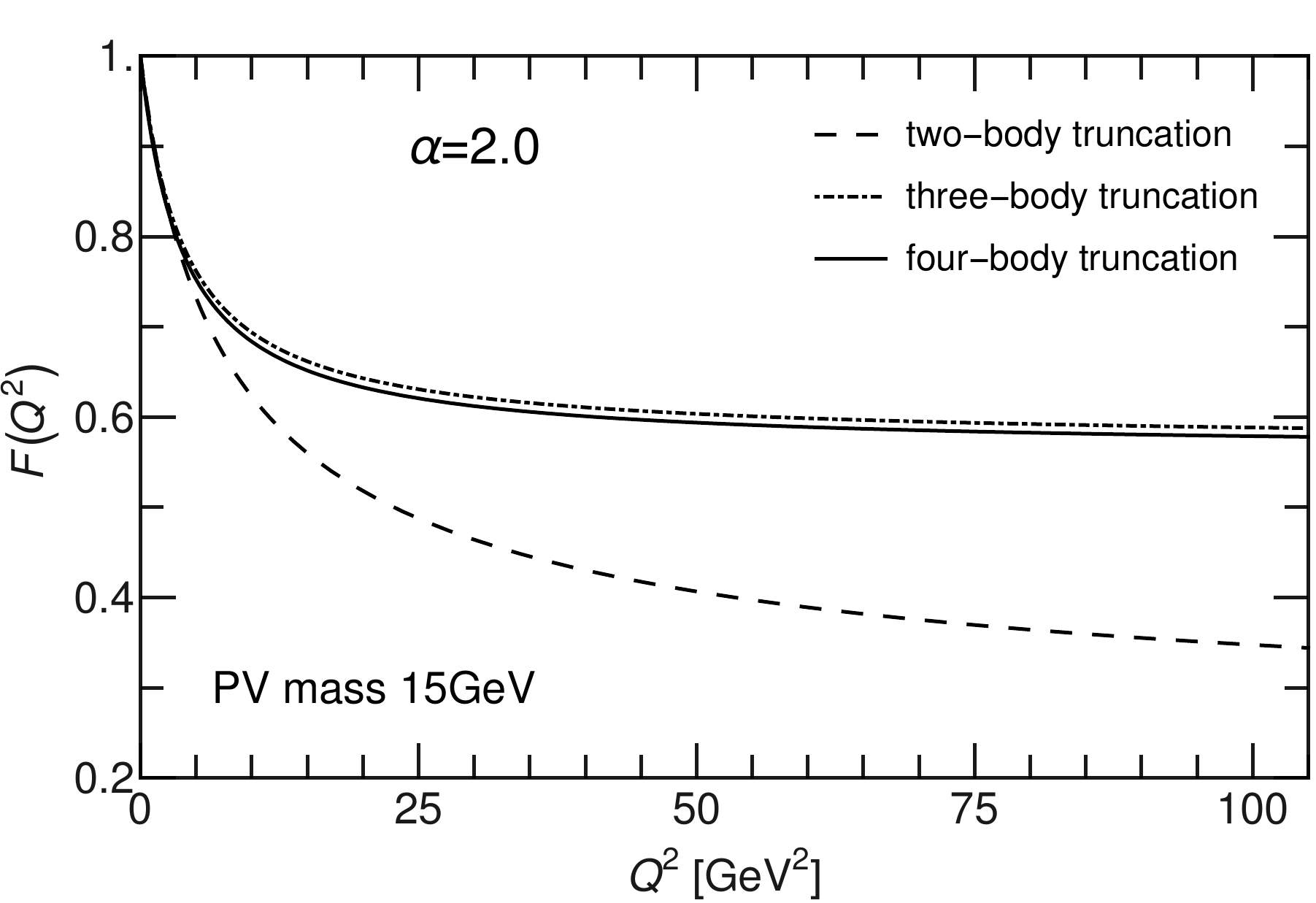}
  \caption{ Comparison of the form factors calculated in the two-,
    three and four-body truncations at $\alpha = 1.0$ (top) and
    $\alpha = 2.0$ (bottom).  The three- and four-body form factors
    are fitted by Eq.~({\ref{eqn:formfactorfit}}).  }
\label{fig:formfactor_conv}
\end{figure}
Figure \ref{fig:formfactor_conv} compares the form factors obtained
from the two-, three- and four-body truncations for two selected
couplings. The three- and four-body truncation results show good
agreement even at the non-perturbative couplings, suggesting a
reasonable convergence with respect to the Fock sector expansion.

\section{Discussion and Conclusions\label{sec 4}}

We solve the single-nucleon sector of the scalar Yukawa model in light-front dynamics
within a four-body (up to one scalar nucleon and three scalar pions) Fock sector
truncation.  Fock sector dependent renormalization is implemented. The coupled 
system of linear integral equations is derived and solved numerically.
The numerical study of the Fock sector norms suggests that up to
$\alpha \approx 1.7$ the system is dominated by the lowest Fock
sectors.
By comparing the form factors from successive Fock sector truncations
(two-, three- and four-body), we find that the Fock space expansion of
the form factor for the scalar nucleon converges as the number of 
scalar pions increases even in the non-perturbative region.

Solving the one-nucleon sector is also the first step for the study of
the two-nucleon sector -- a bound-state problem, which has been
extensively studied in various approaches (see, e.g., \cite{Hiller1993.4647} and the 
references therein).  However not all these approaches are from
first principles.
In our approach, the two-nucleon sector obeys similar integral
equations.  The bare couplings and the mass counterterms, according to
FSDR, are already provided by the one-nucleon sector (up to three
dressing pions).  Therefore, our approach allows a systematic study of
the theory with a non-perturbative renormalization.

This calculation demonstrates that the light-front Tamm-Dancoff,
equipped with the Fock sector dependent renormalization, is a general
\textit{ab initio} non-perturbative approach to quantum field
theories. While the solution of the scalar Yukawa model may be useful
for, e.g., chiral effective field theory studies, this approach has
also been applied to more realistic field theories, including the
Yukawa model (truncation up to one spinor and two scalars)
\cite{Karmanov2012.085006} and QED (truncation up to one electron and
two photons) \cite{Hiller1998.016006}.  
In these theories, the vertex functions also diverge, in contrast
to the scalar Yukawa model. However, after the renormalization, the 
physical observables converge as expected.
Nevertheless, the study of the higher Fock
sector expansion in these models is in principle similar to the
current one, which indicates the potential of this approach as an
alternative to other first-principle methods, e.g. the lattice gauge
theory, especially in the study of hadron structures.

\section*{Acknowledgements}
We are indebted to A.~V.~Smirnov for kindly providing us some
numerical benchmark results for the three-body truncation.
We wish to thank J.~Carbonell, J.-F.~Mathiot and X.~Zhao for valuable
discussions. One of us (V.A.K.) is sincerely grateful to the Nuclear
Theory Group at Iowa State University for kind hospitality during his
visits.
This work was supported in part by the Department of Energy under
Grant Nos. DE-FG02-87ER40371 and DESC0008485 (SciDAC-3/NUCLEI) and by
the National Science Foundation under Grant No.  PHY-0904782.
Computational resources were provided by the National Energy Research
Supercomputer Center (NERSC), which is supported by the Office of
Science of the U.S. Department of Energy under Contract
No. DE-AC02-05CH11231.

%

\end{document}